\documentclass{ws-rv9x6}
\usepackage{graphicx,float}
\usepackage{ws-rv-thm}   
\usepackage{ws-rv-van}   
\makeindex

\makeindex
\begin{document}

\chapter[Phase Transitions and Criticality in Collective Behavior]{
Phase Transitions and Criticality in the Collective Behavior of Animals - Self-organization and biological function
}

\author{Pawel Romanczuk}
\address{Institute for Theoretical Biology, Department of Biology, Humboldt Universität zu Berlin \\
Bernstein Center for Computational Neuroscience, Berlin, 10099 Berlin, Germany \\
Research Cluster of Excellence ``Science of Intelligence'', 10587 Berlin, Germany 
}
\author[P. Romanczuk \& B.C. Daniels]{Bryan C. Daniels}
\address{School of Complex Adaptive Systems, College of Global Futures, Arizona State University, Tempe, AZ 85287, USA.}

\begin{abstract}
Collective behaviors exhibited by animal groups, such as fish schools, bird flocks, or insect swarms are fascinating examples of self-organization in biology.  Concepts and methods from statistical physics have been used to argue theoretically about the potential consequences of collective effects in such living systems. In particular, it has been proposed that such collective systems should operate close to a phase transition, specifically a (pseudo-)critical point, in order to optimize their capability for collective computation. In this chapter, we will first review relevant phase transitions exhibited by animal collectives, pointing out the difficulties of applying concepts from statistical physics to biological systems. Then we will discuss the current state of research on the ``criticality hypothesis'', including methods for how to measure distance from criticality and specific functional consequences for animal groups operating near a phase transition.  We will highlight the emerging view that de-emphasizes the optimality of being exactly at a critical point and instead explores the potential benefits of living systems being able to tune to an optimal distance from criticality.   We will close by laying out future challenges for studying collective behavior at the interface of physics and biology. 
\end{abstract}
\tableofcontents

\body
\section{Introduction}
Collective behavior\index{collective behavior} exhibited by large animal aggregations such as swarms of insects\index{insect swarms}, schools of fish\index{fish schools}, and flocks of birds\index{bird flocks} are ubiquitous and fascinating examples of biological self-organization\index{self-organization}. Over the past decades it has attracted scientists from a wide range of scientific disciplines far beyond biology. Theoretical physicists in particular investigate the analogy between large animal collectives and systems studied in statistical physics, such as fluids or magnets, where local interactions between a large number of rather simple components lead to emergence of novel properties at the macroscopic level, which can be difficult to trace back to properties of the individual components\cite{boccara2010modeling,newman2011complex,kwapien2012physical}. 

In the past century, since the introduction of the famous Ising model\cite{ising1924beitrag,newell1953theory,ising2017fate}. statistical physics has developed a powerful methodology to study self-organization in complex systems and the emergence of order, through the theory of phase transitions\index{phase transitions} and critical phenomena\cite{stanley1971phase,domb2001phase}. Whereas most of the research was devoted to understanding systems at thermal equilibrium, where an encompassing body of theory confirmed by experiments has been firmly established, we have also witnessed decades of research into self-organization and pattern formation in non-equilibrium systems \cite{domb2001phase,hinrichsen2006non}, relevant to biological systems in general, and collective animal behavior in particular.  Whereas many questions regarding non-equilibrium phase transitions remain open, it is beyond doubt that physics provides a diverse and powerful tool set that can further our understanding of the complexity of the living world.   

However, a question which physics is not well suited to address in biology is the ``Why?'', the question on the ultimate causes and the biological function of a particular self-organized behavior. This issue must be viewed and addressed through the prism of evolutionary theory and behavioral biology, and it demonstrates the need for a truly interdisciplinary exchange between physics and biology in understanding complex living systems. 

An intriguing property of living systems that distinguishes them from most (or all) inanimate matter is their ability to react adaptively to changing environments. In general, this capacity for enacting functional adaptive behavior relies on distributed processing of information at various levels, characterized by the collective dynamics of a large number of interacting components or agents constituting the complex biological system. Examples range in scale from regulatory interactions among proteins in cells\cite{davidson2006gene} to neurons interacting in brains\cite{beggs2012being,kinouchi2006optimal} to the group behavior of animals and humans \cite{moussaid2009collective}. 

The parameter space in which such collective biological systems operate is vast, both due to the complexity of single individuals and to the large number of individuals that can constitute a functional group.
The important question emerges whether particular parameter combinations, or parameter regions, are particularly suited for their biological function by rendering their collective behavior in some sense near optimal. In this context, the so-called \emph{criticality hypothesis}\index{criticality hypothesis} has been proposed. It suggests that complex living systems processing information in a distributed way should operate at or close to a critical point that separates qualitatively different aggregate behavior.\footnote{
    To be more precise, a finite-size biological system would be hypothesized to be close to pseudo-critical point: strictly speaking phase transitions and criticality are only properly defined in the thermodynamic limit, in which the number of components $N \rightarrow \infty$.} 
At such critical points, statistical physics predicts maximal susceptibility, i.e.\ maximal sensitivity, of the collective dynamics at the macroscopic level to small differences in an external input, as well as fast transmission of information across arbitrarily large scales\cite{munoz2018colloquium}. 

One challenge for the criticality hypothesis\index{criticality hypothesis} lies in explaining how distributed systems are able to regulate their behavior to stay in the critical region of parameter space.
Research on self-organized criticality\index{self-organized criticality}, a concept introduced by Bak et al in [\refcite{bak1987self,bak1988self}], suggests one set of mechanisms for how complex systems might tune  towards critical points in a self-organized way without external control, which assumes a time-scale separation between the fast (relaxation) dynamics of the system and a slow driving of the system towards the critical point.  In this context, it was also suggested that (self-organized) critical dynamics can explain the overabundance of power-law distributions in empirical data\cite{bak1987self,markovic2014power}. Self-organized criticality experienced significant scientific interest in the 1990s, with the corresponding research focusing mostly on idealized mathematical models\cite{jensen1998self,turcotte1999self}. Nevertheless, the concept remained controversial due to many questions remaining open and, more importantly, its unclear relevance to real-world phenomena. 

After moving out of the scientific focus towards the end of the 20th century, the criticality hypothesis received renewed attention largely based on new experimental observations in the field of neuroscience, following the report of critical neuronal avalanches by Beggs \& Plenz in 2003 \cite{beggs2003neuronal}. Since then, a broad range of theoretical and experimental work has been published on the criticality hypothesis in different biological systems, including neuronal systems\cite{beggs2012being,hesse2014self,priesemann2014spike,plenz2021self}, gene-regulatory networks\cite{balleza2008critical,daniels2018criticality}, and collective dynamics of animal groups\cite{mora2011biological,calovi2015collective,klamser2021collective,poel2022subcritical}. This more recent research is characterized by 1) being more directly connected to experimental observations, 2) suggesting empirically motivated mechanisms for self-tuning towards the critical point, and 3) questioning the sole focus on benefits of criticality, instead emphasizing the importance of adaptively managing competing trade-offs, potentially by actively tuning the distance to critical transitions depending on  environmental context.

A side note on \emph{self-organization}\index{self-organization}: Throughout this chapter we will use this term when referring to the behavior of animal groups. It is a term widely used in the context of complex systems, yet its exact meaning and relevance in the context of biological systems is far from obvious. In physics, self-organization typically refers to spatio-temporal structure formation in systems consisting of rather simple, homogeneous components such as atoms or molecules. Yet animal groups are composed of individuals with a complex behavioral repertoire, and may exhibit high levels of heterogeneity and hierarchical organization. Interestingly, even clonal fish exhibit significant behavioral individuality despite near-identical rearing conditions\cite{bierbach2017behavioural}. If we consider an extreme example of a single dominant individual --- a ``leader'' --- persistently determining the behavior of an entire group, one would not view this as self-organized collective behavior. However, this extreme hypothetical case is rarely encountered in nature. Even for strongly hierarchical societies of eusocial insects, such as bees\cite{wild2021social} or ants\cite{bonabeau1997self,gordon2019ecology}, the colony level behavior is far from being determined by the queen alone or a particular caste within the colony. Furthermore, it has been shown that despite strong rank hierarchies in baboon troops, the actual movement initiation decisions are very much driven by rank-independent consensus mechanisms\cite{strandburg2015shared}. Here, we will use the term \emph{self-organization} in a general sense to refer to the emergence of macroscopic (group-wide) collective behaviors from interactions between individuals, which lead to formation of spatio-temporal dynamics and patterns that cannot be attributed solely to the behavior of isolated individuals or to environmental factors.

In this chapter, we will refrain from a general discussion of self-organized criticality\index{self-organized criticality} in biology, but instead focus specifically on 
the role of criticality in the collective behavior\index{collective behavior} of animals, which has received comparatively little attention. However, when useful we will touch upon selected results in other fields, referring to results in neuroscience relevant to distributed information processing, and comparing animal groups to other biological systems with respect to the criticality hypothesis. We refer readers interested in the broader perspective to some excellent recent review articles on the criticality hypothesis \cite{munoz2018colloquium} in particular in the field of neuroscience \cite{beggs2012being,hesse2014self,plenz2021self}.  

In the following, we will first discuss in section \ref{phasetransitions} the applicability of the theory of phase transitions to the collective behavior of animals, and review the most relevant types of phase transitions in this context. We will continue in with a discussion of methods for quantifying criticality in biological collectives (section \ref{quant}) and ways that criticality can impact biological function (section \ref{function}),  before turning our attention in section \ref{mechanism} to potential mechanisms that may enable animal collectives to tune themselves in a self-organized manner towards or away from critical points. Finally, we will close this chapter with a section on open questions and future challenges regarding the criticality hypothesis in animal groups.

\section{\label{phasetransitions} Phase transitions\index{phase transitions} in collective behavior}


\subsection{The applicability of phase transition concepts to biology}

Biological systems, and animal groups in particular, share some basic properties with classical statistical physics systems in which the concept of a phase transition was first developed.
Yet there are also important differences, and thus the application of the phase transition concept to collective animal behavior needs to be critically assessed. 

First, a crucial difference of biological systems in comparison to systems classically studied in statistical physics is their far-from-equilibrium nature. The theory of phase transitions under non-equilibrium conditions is a very active field of research of modern statistical physics \cite{marro2005nonequilibrium,hinrichsen2006non,henkel2008non}. While many questions remain open, there is no fundamental reason to believe that the corresponding theoretical concepts do not extend to living systems, including animal groups, which is further supported by a large body of literature on phase transitions in biology over the past decades \cite{vicsek1995novel,BuhSumCou06,szabo2006phase,FeiPinGel18}.    

Second, an important difference is the sheer size of the systems in terms of the number of constituting units. Whereas a macroscopic volume of matter typically contains $N\sim10^{23}$ particles, a typical animal aggregation consists of $N\sim 10-10^3$ individuals. In rare cases, significantly larger collectives with $\sim 10^6$ individuals are observed, typically in the context of large migration movements in animals such as pelagic fish (e.g. sardines) \cite{misund2003schooling} or desert locusts \cite{BuhSumCou06}. 
However, phase transitions in statistical physics are rigorously defined only in the thermodynamic limit of extremely large, or more precisely infinite, systems. The mathematical abstraction of $N\to\infty$, while a very good approximation for the bulk behavior of ``classical'' physical matter, is certainly questionable for describing most biological systems and animal groups in particular. 
This critique, while correct, does not refute the significance of phase transitions and criticality in the description of finite-sized biological systems. The conjectured benefits of (quasi-)criticality for collective animal behavior do not rely on the assumption of the thermodynamic limit. Various aspects of collective information processing, such as correlation lengths, information transmission, and susceptibility to inputs, still become maximal at quasi-critical points in finite-sized systems \cite{kinouchi2006optimal,calovi2015collective,munoz2018colloquium,poel2022subcritical}. 

However, the theoretical concept of universality\index{universality} is intrinsically linked to the concept of thermodynamic limit through the application of renormalization methods.  Universality predicts the emergence of scaling laws and critical exponents that depend only on fundamental properties of the system such as dimensionality and symmetry, and not on microscopic details.  This simplified scaling behavior may only be observed above a critical system size. Unfortunately, due to the complexity of social behavior of animals, there are massive gaps in our knowledge on the nature of social interactions and fluctuations. Therefore even rough estimates of corresponding critical system sizes, above which universal behavior in the statistical physics sense could be observed, appear currently impossible to be established. Thus, it is imperative to be extremely cautious when interpreting empirically observed scaling laws in small to mesoscale animal groups in the context of universality classes. 

For these reasons, it is our opinion that categorizing biological phase transitions into distinct universality classes is less practical than in classic systems from physics, and may indeed be impossible.  Still, the concept that a single relevant combination of parameters dominates collective behavior near a transition, and that this might allow for immense simplification of effective models of particular systems, remains viable.

Due to the relatively small system sizes, boundary conditions must be expected to play a non-negligible --- if not a dominating --- role for self-organized collective behavior\cite{cavagna2013boundary}. For many statistical physics theories, this would be considered as rather problematic. However, from a biological point of view this is likely an important or even a defining feature of animal collectives \cite{davidson2021collective}. If we consider distributed sensing of environmental cues and collective processing of this information as the core function of animal aggregates, for example in the context of predator detection or food search, then it becomes obvious that boundaries are of fundamental importance; e.g.\ in visual perception, most of the environmental information will be perceived by individuals at the edge of the group.

\subsection{Flocking\index{flocking}}
In the context of animal groups, arguably the most easily observed phase transition is the emergence of orientational order due to spontaneous symmetry breaking \cite{BuhSumCou06}.  This flocking transition separates a disordered state, with individuals moving in random directions with a vanishing center of mass speed, from an ordered flocking state, with a non-vanishing average momentum of the entire system.  The phenomenon of flocking is strikingly visible in groups of birds such as starlings \cite{ballerini2008interaction}.  

The first theoretical treatment of flocking\index{flocking} as a phase transition was reported by Vicsek\index{Vicsek model} and co-authors in 1995 in their seminal work on self-propelled particles moving with constant speed $v_0$ and interacting with a ferromagnetic (or polar) alignment interaction\cite{vicsek1995novel}. It immediately received a lot of attention as the reported emergence of long-range orientational order in this non-equilibrium extension of the classical XY-model appeared to violate the Mermin--Wagner\index{Mermin--Wegner theorem} theorem\footnote{In equilibrium systems, the Mermin--Wagner theorem predicts that no long-range orientational order is possible in two dimensions, and the original Vicsek model was formulated in 2D.} \cite{mermin1966absence,hohenberg1967existance}. Very soon after, Toner and Tu were able to show that the non-equilibrium nature of the model for a non-zero self-propulsion speed makes the decisive difference \cite{toner1995long,toner1998flocks,toner2005hydrodynamics} why the Mermin--Wagner theorem does not apply. Following these initial publications, the nature of the transition in the Vicsek model\cite{gregoire2004onset,chate2008modeling,baglietto2009nature,ginelli2016physics}, and more generally in models of self-propelled particles with spatially local alignment interactions, both polar and nematic, has been intensely researched \cite{chate2006simple,peruani2011polar,baskaran2012self,giomi2012banding,bertin2015comparison,grossmann2016mesoscale}. Whereas originally the transition was believed to be continuous\cite{vicsek1995novel,baglietto2009nature}, later systematic numerical simulations as well as theoretical analyses have shown that the homogeneous ordered state is unstable in the vicinity of the critical point with respect to longitudinal density modulations \cite{gregoire2004onset,bertin2006boltzmann,ihle2011kinetic,ihle2013invasion,ginelli2016physics}. This leads to formation of large-scale, high-density bands moving through a disordered, low density gas-like ``background''. The emergence of these spatial heterogeneities eventually changes the nature of the phase transitions to a discontinuous one. However, the discontinuous nature of the transition is often masked by strong finite size effects and can only be reliably observed at very large system sizes and/or high self-propulsion speeds \cite{chate2008collective}. The fundamental mechanism behind the density instability is the presence of a density-order coupling, where, on average, regions with higher density are also more ordered \cite{bertin2009hydrodynamic}.    

There exists a large variety of flocking models of self-propelled particles, some even lacking explicit alignment interactions\cite{szabo2006phase,grossman2008emergence,hanke2013understanding,grossmann2013self}. However, as long as the interactions are short-ranged and result in effective alignment while the system exhibits fluid-like lack of positional order, the transition towards an ordered state will resemble the one observed in Vicsek model, including the above mentioned density-order coupling, and can be assumed to fall into the same universality class, if it can be defined. There exists also a large class of systems which consider self-propelled particle systems with attractive and repulsive forces exhibiting potentially different types of flocking transitions\cite{gregoire2004onset,romanczuk2009collective,ferrante2013elasticity}. 

Based on a detailed analysis of experimental data obtained from 3D tracking of starling flocks, it has been suggested that interactions between pairs of individuals are governed by topological distance rather than the metric distance\cite{ballerini2008interaction}. Such interactions in corresponding topological flocking models can be also long-ranged, as a focal individual pays attention to others if they are within a set of nearest neighbors, independently of their Euclidean distance\cite{ginelli2010relevance,rahmani2021topological}. This metric-free nature of the interaction was assumed to disable the density-order coupling, thus eliminating the density instability and resulting in a continuous flocking transition\cite{ginelli2010relevance}. However, recent research into flocking models with distance-independent $k$-nearest-neighbor interactions has shown the formation of bands due to a weak yet non-vanishing density-order coupling \cite{martin2021fluctuation,rahmani2021topological}, which can be further enhanced by the presence of spatial heterogeneities\cite{rahmani2021topological}.  

Most flocking models assume movement of individuals with a constant speed. This simplifying assumption not only reduces the model complexity but offers a direct analogy to the fixed spin amplitude in closely related statistical physics models like Ising, Potts or XY models. However, animals moving in groups typically exhibit variable speed that can be modulated by social interaction, and thus has to be considered as an additional degree of freedom \cite{grossmann2012active,klamser2021impact,cavagna2022marginal}. Scale-free speed correlations observed in bird flocks can only be explained by variable speed models at criticality\cite{BiaCavGia14}.  It has been also shown that variable speed may dramatically alter and expand the self-organized collective behaviors and lead to new types of order-disorder transitions \cite{grossmann2012active,klamser2021impact}.

Last but not least, recently, based on empirical observations of highly polarized, collective turning behavior in bird flocks, so-called inertial spin models featuring non-dissipative couplings have been proposed\cite{attanasi2014information,cavagna2015flocking}.  In contrast to the dissipative Vicsek model, the inertial spin model has been reported to correctly reproduce the dynamical correlations of velocities and non-exponential relaxation dynamics \cite{cavagna2019dynamical}.

\subsection{Collective decision making}

Collective decision making\index{collective decision making} is another prominent example of collective animal behavior, where symmetry-breaking phase transitions play an important role.
Examples include social insects choosing a new nest site\cite{visscher2007group}, fish schools or baboon troops choosing where to forage \cite{couzin2011uninformed,strandburg2015shared}, and social groups coming to consensus about power hierarchies\cite{DanFlaKra17,BruKraFla18}. 

Collective decision making in biology can take place across a variety of spatial and temporal scales, but the fundamental decision dynamics can often be well captured by simple, non-spatial models\cite{arganda2012common,pais2013mechanism}.  
Most collective decision models consider binary decision tasks, which can be modelled by Ising-type models \cite{turalska2009complexity,hartnett2016heterogeneous,pinkoviezky2018collective}, while multi-choice decisions can in principle be described with a Potts model \cite{lee2014simple}. However, the majority of collective decision making models considers some sort of threshold or quorum interactions, where an agent instantaneously updates its decision state based on the absolute number, or relative ratio, of its neighbors already committed to a certain choice. Thus instead of formulating the model in terms of a Hamiltonian as a starting point, these models are typically formulated in terms of behavioral algorithms in discrete time. An encompassing review of a variety of such models is given in [\refcite{castellano2009statistical}]. Another complementary approach of modelling collective decisions is rooted in dynamical and stochastic system theory and involves formulation of (stochastic) differential equations or the evolution of continuous decision variables of interacting agents\cite{srivastava2014collective,tump2020wise}. 

In general, a collective decision is defined as the commitment of a majority of a group to a single option, starting from an initially undecided collective state. Thus it directly corresponds to a breaking of symmetry. A major difference from flocking literature is that models of collective decision making typically assume either well-mixed or all-to-all interaction between agents\cite{arganda2012common,pais2013mechanism}, or a static interaction network ranging from lattices \cite{turalska2009complexity,hartnett2016heterogeneous} to scale-free\cite{gronlund2007dynamic} and small-world networks \cite{winklmayr2020wisdom}. Notable exceptions are works combining explicit spatial dynamics, such as flocking, with collective decision making \cite{couzin2005effective,couzin2011uninformed}, which allow for study of the interplay of spatial self-organization of the group structure and the decision dynamics. Here, a fascinating example of spatially-explicit collective behavior is the cooperative cargo transport by ants \cite{GelPinFon15,FeiPinGel18}.  

Interestingly, despite the obvious analogy to phase transitions, the discussion of criticality in the context of collective decision making\index{collective decision making} has received little attention (see e.g.\cite{GelPinFon15,DanRom21} ). This may be related to the strong interdisciplinary nature of the field, as well as the rather small group sizes studied. Furthermore, modeling of collective decision making is often rooted in the theory of dynamical systems and stochastic processes, with transitions between states often discussed in the context of bifurcation theory. 
While there are important conceptual differences between bifurcations and phase transitions\cite{bose2019bifurcation}\footnote{The main difference is that phase transitions rely on a thermodynamic limit in which the number of components goes to infinity, while bifurcation theories rely on a steady-state limit in which time goes to infinity.}, the functional benefits of criticality, such as maximum sensitivity to external perturbations, are also a feature of bifurcations in dynamical systems\cite{gross2021not}. 
Finally, rather than arguing for a single optimal point in parameter space as in the criticality hypothesis, the collective decision making literature focuses on functional trade-offs, such as the ubiquitous speed-accuracy trade-off. Interestingly, speed and accuracy of collective decisions are strongly modulated by the distance from a critical point, as we have recently shown by investigating collective decision making on networks \cite{DanRom21}.

\subsection{Behavioral contagion}\index{behavioral contagion}

Another commonly observed pattern in collective behavior is the spread of behavior or information through a group, resulting in behavioral cascades that can quickly encompass the entire collective \cite{dodds2004universal,bottcher2017critical,poel2022subcritical}. Examples include startles spreading through a fish school \cite{sosna2019individual} and conflict spreading through a macaque society \cite{DanKraFla17}.

In contrast to flocking and collective decision making, this type of collective dynamics typically does not correspond to a spontaneous symmetry-breaking transition but rather to percolation \cite{araujo2014recent}, and is directly related to the non-equilibrium phase transition studied in epidemic spreading. Depending on a control parameter, e.g.\ coupling strength, a behavior initially adopted by a few individuals may stay localized to a small neighborhood, or spread quickly in an avalanche-like manner through the entire group, similar to a pathogen. One distinguishes two fundamental types of spreading processes: simple and complex contagions. In a simple contagion, the ``infection'' of a focal individual results from a superposition of independent pair-wise contacts with active or infected neighbors. In contrast, in complex contagion the infection probability depends the state of the entire neighborhood, i.e.\ on higher-order interactions, and cannot be decomposed into binary interactions.  Disease spreading represents a classic example of simple contagion, whereas spreading of behavior is in general assumed to be a complex contagion processes \cite{granovetter1978threshold,rosenthal2015revealing}. In particular, various threshold models proposed for social contagion represent examples of complex contagions \cite{granovetter1978threshold,ruan2015kinetics}.

In the well-mixed case, simple contagion processes exhibit a continuous phase transition between a phase of vanishing infection load to a phase with a finite fraction of the system being infected. Complex contagion, on the other hand, may exhibit different types of transitions depending on model details, even at the level of mean field analysis \cite{araujo2014recent,dodds2004universal, bottcher2017critical,iacopini2019simplicial}. In 2004, Dodds \& Watts proposed a generalized model of contagion \cite{dodds2004universal,dodds2005generalized}. The model is formulated in discrete time and the coupling between individuals is determined by the probability $p$ of sending an activation signal ( ``infection dose'') to a neighbor, which by definition is bounded between $0$ and $1$. Depending on the model parameters, in particular the distribution of thresholds, Dodds \& Watts distinguished three fundamental types of transitions: (I) a continuous onset of an infected state above a critical interaction probability $p>p_c$ for dose-response behavior resembling simple contagion, (II) a discontinuous one with a bounded coexistence region, and (III) a discontinuous one with a coexistence region that extends over the entire possible infection probability range. We note that, in order to  re-formulate the model in continuous time, one has to replace the probability $p$ by a probability rate $\rho$, which is unbounded, as the probability of infection interaction over a short time interval $dt$ is given by $p_{dt}=\rho dt$ \cite{sosna2019individual}. Thus, it is unclear whether the two discontinuous transition types (II and III) are indeed fundamentally different or just a consequence of the model formulation. 

Recently, B\"ottcher {et al}\cite{bottcher2017critical} have analyzed analytically the mean-field behavior of a related contagion model formulated in continuous time and confirmed the existence of both continuous and discontinuous transitions.  

\subsection{Synchronization}\index{synchronization}

When individuals in a group engage in periodic behaviors,
interactions can cause their behavior to become synchronized.
One human example is the synchronization of clapping
in an audience, and in biology synchronization is observed
in the collective activity of neurons, in the behavior of
heart cells, and in the flashing 
of certain species of fireflies \cite{SarHayPel21}.

One classic model for synchronization is the Kuramoto 
model \cite{Kur84,AceBonVic05}\index{Kuramoto model}, in which oscillators influence 
one another via pairwise 
coupling that depends on the difference between their
phases.  Given variation in the natural frequencies for
interacting oscillators, 
a continuous phase transition controlled by the interaction strength
separates unsynchronized activity from a collective state 
in which a fraction of the oscillators oscillate at 
a single consensus frequency \cite{Kur84}.  Other model 
variations, for instance oscillators that interact only through
momentary pulses \cite{MirStr90} 
also show similar phenomenology of synchronization. 

The language of statistical physics is useful for 
describing the synchronization of large groups of oscillators, 
particularly using the concept of
an order parameter.  For phase synchronization, the relevant order parameter is the amplitude of the mean field, treating
each oscillator's state as a unit vector in the complex plane.

\section{\label{quant} Quantifying criticality in biological collectives}

If we suspect that a living system exists near a collective transition, how can we test this and make the statement more precise?  Typical methods from statistical physics can be useful, though applying to animal groups often requires modification due to the importance of finite size and nonequilibrium dynamics.

One generic strategy for continuous phase transitions is inspired directly by statistical physics: Define a collective order parameter and look for peaks in the sensitivity of the order parameter to external perturbations (also known as susceptibility in the language of magnetic systems).  A peak in sensitivity corresponding to a collective instability becomes a convenient way to define an effective critical point in a finite system \cite{DanEllKra16,DanKraFla17,poel2022subcritical}.  This approach is often a natural one because the sensitivity of global behavior to changes at the individual scale 
is commonly connected to
hypothesized functions of criticality.  More generally, order parameters
can be defined in terms of a Fisher information measure that acts as a generalized 
sensitivity \cite{ProLizObs11}.
Finally, in spatial cases, the correlation length can
similarly be measured and compared to the peak expected
at criticality.

Phase transitions in animal groups often involve cascades of activity, where the relevant transition is 
the percolation-like transition between cascades that quickly die out and those that spread through
the entire system.
In this case, more specific measures of distance from
criticality can be used.  First, a measure
of local amplification: the degree to which
changes to individuals spread to change the state of other
individuals.  In epidemiology, this corresponds to the reproduction number $R_0$, measuring the number of other individuals infected directly due to one individual becoming sick \cite{vynnycky2010introduction}.  In an infinite system, an amplification value of 1 identifies the critical point. In finite systems, this corresponds only to local instability, but can still be useful as a rough indicator for when collective effects will be maximized \cite{poel2022subcritical,DanKraFla17}.
Second, a global measure can be constructed from
cascade size distributions, which are 
expected to decay like power laws at the transition.
Measuring the deviation from a power law has been used as an indicator of criticality (for instance, in neural avalanches \cite{SheYanPet09}). 

Detecting and characterizing discontinuous transitions has generally been less emphasized in the literature.  Yet discontinuous transitions do occur in models of collective behavior.  For instance, in a model of fish schooling with resource detection, evolutionary stable states occur near discontinuous transitions between cohesive and dispersed collective states \cite{HeiRosHag15}.  Such transitions can be detected by measuring bistability in the collective order parameter in the vicinity of the transition, or hysteresis in the presence of time-dependent driving.

Beyond merely demonstrating that a system is ``near'' a
transition,
a main goal for these measures of criticality is to help
reason about how system parameters are tuned.  In particular, near a continuous phase transition, changes to  collective behavior are dominated by changes in the
distance from the transition, with changes to other dimensions in parameter space having relatively small contributions.  
This can simplify the question of how individual-scale
parameters must be tuned to achieve functional collective behavior.

\section{\label{function} Consequences of phase transition phenomena on biological function}

\begin{figure}
\centering
\includegraphics[width=0.75\textwidth]{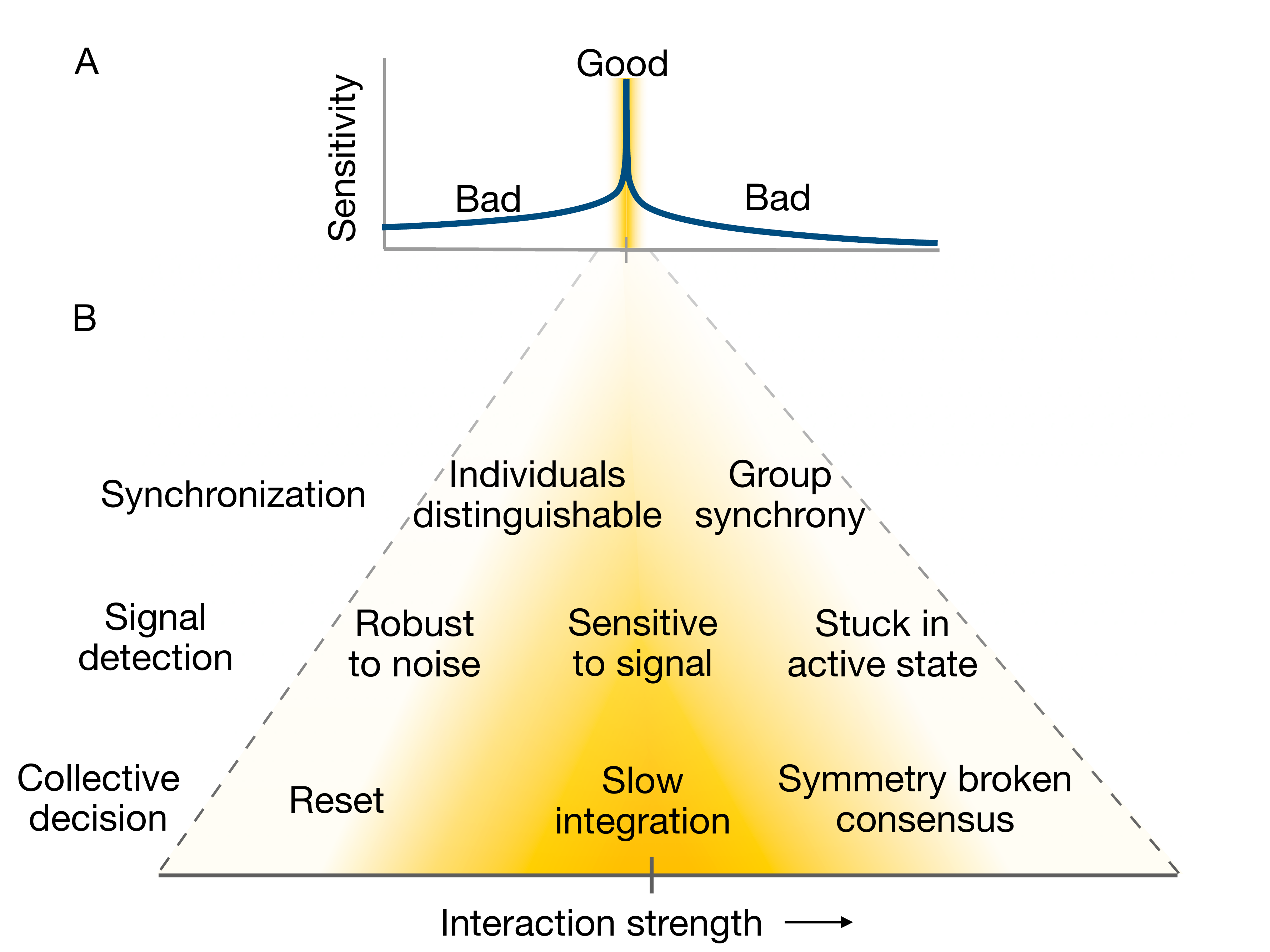}
\caption{Biological function near a continuous phase transition.  (A) A simplistic view recognizes that collective sensitivity is enhanced only near the critical point --- hence the idea that being tuned near criticality can be functionally advantageous.
(B) Sufficiently near the transition, collective effects related to the transition are graded.  Functional consequences, which can be advantageous or harmful depending on context, can be traded off against one another.
\label{fig:schematic}%
}
\end{figure}

The most basic view of the relationship between collective transitions and biological function is
summed up by the fact that, when a system is far away from a transition,
its collective behavior is boring (Fig.~\ref{fig:schematic}A).  
For instance, when interactions are very weak, then the system's responsiveness is unenhanced by 
the group, simply given by the sum of the responsiveness of each component.  
When interactions are very strong,
the system is likewise unresponsive because it is difficult to modify the system's 
prevailing order (or all order may be lost in the case of a chaotic system).  
It is only near a collective transition that information exchange among individual 
components can hope to have an appreciable effect on the macroscopic behavior.
This idea has been explored in a number of theoretical treatments that identify parameter regions of marginal stability --- the ``edge of chaos'' --- as defining dynamics that are capable of nontrivial information exchange and computation \cite{langton1990computation,teuscher2022revisiting}.

Beyond this basic assessment, there are a number of distinct phenomena that happen near collective transitions that can be functionally
relevant. Yet these collective phenomena are not necessarily beneficial. 
This fact, along with the typical ``blurriness'' of biological transitions arising from finiteness and heterogeneity, motivates zooming in near collective transitions to study how varying slightly away from maximal sensitivity affects biological function (Fig.~\ref{fig:schematic}B). 
Going beyond the simple statement that ``criticality is best'', this more nuanced perspective often leads to trade-offs that can be managed by the system maintaining a distance from the transition or even actively changing this distance\cite{poel2022subcritical,cramer2020control}.  

\begin{figure}
\centering
\includegraphics[width=0.95\textwidth]{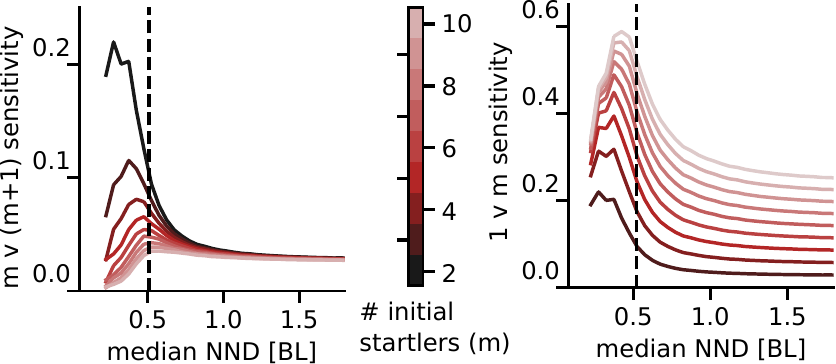}
    \caption{Maximal susceptibility at criticality in a behavioral contagion model as studied by Poel et al.\cite{poel2022subcritical}: The sensitivity, defined via the difference in the average relative cascade sizes triggered by different numbers of initially activated individuals, $m$ vs $m+1$ (left) and $1$ vs $m$ (right), is shown as a function of the median nearest neighbor distance (NND). The NND is a control parameter determining the coupling between individuals and the probability of an activation to spread. This susceptibility measure becomes maximal in the vicinity of the quasi-critical point, which separates a phase of only small-scale, local cascades (high NND, weak coupling) from a phase of global activation (low NND, strong coupling). In functional terms, the system is best at discriminating different numbers of initially active agents at criticality. The vertical dashed lines indicates an analytical approximation of the critical point. Figure taken from Supplementary Information of Poel et al.\cite{poel2022subcritical}
\label{fig:susceptibility}%
}
\end{figure}

Continuous transitions (``critical points'')\index{critical point} in particular have been 
the subject of much speculation regarding impacts on biological function\cite{mora2011biological,munoz2018colloquium}.  
An oft-cited advantage of criticality is 
maximal susceptibility of the aggregate state.
That is, at a critical point, 
changes to behavior of individual animals in a group 
are amplified to have maximal control over the 
group's behavior (Fig.~\ref{fig:susceptibility}) \cite{vanni2011criticality,calovi2015collective,GelPinFon15}.  This can clearly be beneficial when individuals have relevant and accurate information to share, but can equally well be detrimental, as when individual information is noisy or individuals' motives are not aligned. For this reason, being near but not directly at a collective transition has been
hypothesized to be important for maintaining (and perhaps actively tuning) a particular
tradeoff between robustness and sensitivity.  This is thought to be the case in fish
schools responding to a threatening stimulus, where the optimal spreading of startle
behavior can depend on the current threat of predation \cite{poel2022subcritical}.  The fastest spread of
information happens at the transition point, but this can be suboptimal if false alarms 
are much more more likely (and therefore on average more costly) in an environment
with low predation risk.
Such tuning between robustness and sensitivity 
has also been hypothesized
to be important for understanding conflict spreading in 
macaque societies.  Here, individuals have both noisy information 
and competing interests regarding the collective social hierarchy, 
but can nonetheless achieve beneficial consensus \cite{BruKraFla18}
perhaps by actively controlling the distance from a 
transition point where conflict outbreaks are maximally 
sensitive to the behavior of individuals \cite{DanKraFla17}.

A second potential advantage of a continuous transition is the dynamical
phenomenon of critical slowing down\index{critical slowing down}. In neuroscience, 
critical slowing is hypothesized to be a mechanism through which 
long timescales can be created using components with short 
memories \cite{DanFlaKra17}.  Similarly, in decision-making
in animal groups, slow timescales may be important for 
integrating over noisy information \cite{DanRom21}.
In situations with small signal-to-noise ratio, 
this phenomenology also implies a trade-off between speed and
accuracy as the distance from the transition varies, since slower
dynamics allows for more time to integrate noisy information.
The speed--accuracy trade-off in collective 
decisions is thought to be important for many types of 
animal groups, and has been studied extensively in 
social insects \cite{franks2003speed,marshall2009optimal}.  

Collective transitions in general can display
the phenomenon of spontaneous symmetry breaking\index{spontaneous symmetry breaking}, 
in which aggregate states
select a particular global order even when multiple 
possible orderings are equally likely.  In an animal group,
coming to a single group consensus about a decision may be
important even if all possible decisions are equally 
favored, in which case spontaneous symmetry breaking
could be advantageous.  On the other hand, a system that 
remains at a
continuous transition would have large, slow fluctuations that 
explore the alternate global symmetries, and moving 
beyond the critical point could be required to force long-term commitment or 
consensus on a particular collective decision \cite{AreJinDan18,DanRom21}.

While much attention has focused on the theoretically rich area of continuous transitions, discontinuous transitions can be just as important in biology.\footnote{Note that, given the typical blurriness and nonequilibrium nature of biological transitions, it may be difficult to resolve whether a given transition is continuous or discontinuous --- in our view, this categorization is not always important in analyzing a particular transition's functional consequences.}   In particular, hysteresis can create a sustained switch-like 
response.  This can be advantageous in collective decision-making when consensus or strong commitment is important.  For instance, a bee swarm committing to a particular nest site may use strong hysteresis to avoid splitting its individuals between two similar sites.  On the other hand, hysteresis can be costly when it leads to being trapped in a suboptimal collective state. 

Finally, while synchronization is observed across many
animal groups, its functional relevance is still debated.  
Generally speaking, it is likely that multiple 
costs and benefits of coordination
must be weighed against those of acting alone.
In fireflies, not all species synchronize, and in those that do,
they tend to synchronize only at high density, suggesting that
synchronization is either difficult or not always beneficial.
The selective advantage is
not fully understood but is likely to involve 
both female choice of mate and avoidance of predators \cite{LewCra08}.

\section{\label{mechanism} Mechanisms of self-organization towards criticality}

A crucial open question with respect to the criticality hypothesis in animal collectives is that of the actual mechanisms that enable animal groups to tune towards critical points in a fully distributed manner. Here, the main challenge is that phase transitions, and therefore critical points, are macroscopic properties of the system at the level of the entire group, while the adaptation of behavior can be only performed by individuals. First, in the common case that single individuals can only perceive their local neighborhood, it is unclear how they can know at all the global state of the group, and in consequence how they are able adapt their behavior to tune the entire collective towards or away from a phase transition. Second, it has also to be assumed that at any time only a fraction of individuals may adapt its behavior, which in general will increase the heterogeneity of the group with unclear consequences on the ability of the system to tune the macroscopic state, or potentially even detrimental effects on collective behavior by affecting the spatial structure of the group up to a loss of cohesion. 

Regarding the first point of individuals having access only to local information, a potential solution has been proposed in the field of neuroscience, as a mechanism for self-tuning of neuronal networks towards criticality based only on local node dynamics \cite{bornholdt2000topological}. The basic idea is that local averages over time provide some information about the global state of the network. The network can evolve towards a global critical point when individual neurons that are too silent or too active over time decrease or increase their response thresholds, respectively. This mechanism relies implicitly on two important properties of the network dynamics: (1) ergodicity, i.e.\ the assumption that the temporal average of local dynamics corresponds to an ensemble average over the entire network, and (2) time-scale separation between the fast firing dynamics of individual neurons and the slow adaptation dynamics of individual nodes changing their thresholds or synaptic weights. Both assumptions are likely violated in animal groups, leading to open questions about the applicability of the mechanism to self-tuning towards criticality in animal collectives.

In general, while the benefits of criticality for collective information processing have been addressed in various studies, the actual mechanisms on how animal collectives could tune themselves towards critical points have been rarely investigated \cite{attanasi2014finite,hidalgo2014information,brush2016content,klamser2021collective}. 
Here, different adaptive mechanisms are potentially thinkable.

First, for uncovering the proximate mechanisms of self-organization towards criticality, it is important to identify how changes in individual behavior modulate macroscopic group properties, and how these may control the distance to criticality at the collective level. A fundamental variable at the macroscopic level is $N$, the number of individuals in the collective. In finite-sized systems the location of the maximum of the susceptibility is a function of $N$ due to finite size scaling, and it has been argued that in midge swarms it may be the decisive parameter tuning the system in a self-organized way towards criticality\cite{attanasi2014finite}. 

For a fixed group size, group density plays an important role. Assuming that influence of a neighbor on a focal individual is distance dependent, a change in density will directly impact the average strength of social interactions in a group. In fact, it has been recently shown that density modulation is the prime mechanism for tuning the collective escape responses in fish due to changes in perceived risk of predation \cite{sosna2019individual}. Surprisingly, adaptations of individual response thresholds appear to play a negligible role in this context \cite{sosna2019individual}. However, in other ecological contexts or different animal species, adaptation of individual sensitivity to social cues (analogously to synaptic plasticity in neurons), while keeping the typical nearest neighbor distance unchanged, is a reasonable and biologically viable individual-level mechanism for tuning the distance to criticality. 

Finally, individuals can exhibit high levels of heterogeneity in their behavioral traits, and may even assume very different roles in a collective (``\emph{division of labor}''). Therefore changes in the composition of the group at similar group size and density may provide another indirect mechanism for self-organization toward or away from a critical point. Here, however, one should note that strong heterogeneity could lead to fundamental changes in the phase diagram of the system, for example by facilitating the emergence of  Griffith phases, extended regions in parameter space of (quasi)-critical behavior \cite{moretti2013griffiths}.

Over longer time scales, evolutionary adaptations\cite{hidalgo2014information
,klamser2021collective} can tune the behavioral parameters of individuals towards critical points of the collective behavior. This would first require that being critical is on average beneficial across all ecological contexts encountered by animal collectives. Secondly, the critical point --- if beneficial for collective computations --- represents a group level optimum, while evolutionary selection is assumed to act predominantly at the level of individuals. 
In general, there is no guarantee that group optima correspond also to evolutionary stable states with respect to individual-level adaptation. On the contrary, so-called social dilemmas, situations where evolutionarily stable individual strategies differ from group-level optima are rather the rule than the exception\cite{klamser2021collective}. Multi-level selection \cite{o2008good} could provide a mechanism for evolutionary adaptation towards group-level optima. However, recent mathematical analysis demonstrated that as soon as individual-level selection plays a non-negligible role --- which can be safely assumed for most, if not all, animal systems --- group-level optima in general differ from evolutionarily stable strategies \cite{cooney2019replicator}. 

Interestingly, Hidalgo et al\cite{hidalgo2014information} have shown that agents modelled as boolean genetic networks do evolutionarily adapt toward  phase transitions if exposed to heterogeneous, variable environments. Further, they have also shown that interacting groups of such agents, with environmental input coming from other members of the group, also self-tune to criticality. However, a recent study by Klamser \& Romanczuk\cite{klamser2021collective} questions the generality of this result with respect to collective behavior of animal groups. It considers a spatially-explicit model of prey agents performing cohesive, collective motion subject to attacks by a predator agent, and investigates both whether being close to the symmetry-breaking flocking transition is optimal for collective predator response and whether individual-level evolution provides a viable mechanism for these systems to tune towards criticality. While being at criticality indeed appears to be beneficial for the entire collective in terms of optimal information propagation and minimal predation risk for individuals in the group, it turns out that the critical point is not an evolutionarily stable state. On the contrary, the phase transition turns out to be maximally evolutionarily unstable, as that is where the collective dynamics is most sensitive to heterogeneity in individual behavior. This results in individuals with different behavioral parameters having the largest possible differences in their ability to avoid predation, and therefore leads to steepest fitness gradients accelerating evolution of individual behavior away from the critical point deep into the ordered flocking state\cite{klamser2021collective}.

A final self-tuning mechanism for animal groups toward critical points could be some sort of (reinforcement) learning on much shorter time scales than evolutionary adaptations. However, a similar question as discussed above is relevant here as well: How is the system able to tune to a global collective state despite predominantly individual-level learning and access to only local information? In the context of multi-agent reinforcement learning, further questions may arise about the convergence of the learning process in a multi-agent setting \cite{bucsoniu2010multi}.

\section{\label{futureChall} Future challenges \& concluding remarks}

The emergence of novel computational techniques for tracking individuals in large animal collectives\cite{romero2019idtracker,walter2021trex} provides the ability to quantify their collective behavior at unprecedented scale, as well as temporal and spatial resolution. This enables tests of previously theoretical predictions and theory-driven hypotheses, including the criticality hypothesis. And indeed, through a combination of experimental and theoretical work, grounded in statistical physics, it is now possible to quantify whether animal collectives operate close to critical points \cite{attanasi2014finite,poel2022subcritical}. Undoubtedly, depending on the particular system and behavior of interest, this task may be technically very challenging, even if the corresponding methodological approaches are known. In particular, the inverse problem of inferring models of collective behavior from data is known to be difficult \cite{Dan20}, with potential pitfalls for spurious detection of criticality \cite{SchNemMeh14}. 

For future research, it appears that the fundamental open questions to be addressed, both from theoretical and empirical point of view, are about the actual mechanisms on how animal collectives can adapt their distance to criticality in a fully distributed manner, and with only local information available to individuals. Here the evolutionary perspective and the existence of potential social dilemmata between group level optima (criticality) and individual-level evolutionary stable behavioral strategies cannot be ignored.   

As laid out in this chapter, phase transition theory and the concept of criticality are highly relevant for understanding and quantifying the interplay of self-organization and function of collective behavior in biology. The ``criticality hypothesis'', stating that complex biological systems should operate in the vicinity of phase transitions in order to maximize their collective computation capabilities, has the potential to provide a unifying principle potentially relevant to a wide range of different biological systems. It sets up a theory-driven conceptual framework which can inspire, and even structure, entire research agendas at the interface of empirical and theoretical research on collective information processing systems in biophysics and biology. Even in cases where research fails to demonstrate criticality in collective behavior, or even appears to disprove it, exploring the reasons for such ``negative'' results enhances our understanding of collective behavior and generates new research questions and hypotheses \cite{klamser2021collective,poel2022subcritical}. This also very much aligns with recent calls from within biology for more novel ideas grounded in theory \cite{nurse2021biology}. 
However, investigating criticality in biological systems should not be done in epistemic isolation from the physics perspective. On the contrary, we argue that connecting corresponding research to the large body of theories in biology should be a priority, starting from the most fundamental one, the evolutionary theory, to more specific ones such as the many-eyes hypothesis in collective anti-predator behavior \cite{lima1995back} or theories on the role of ecological factors\cite{gordon2014ecology,monk2018ecology,davis2022using}. 
A truly bidirectional exchange between physics and biology opens new avenues of research with enormous potential for better understanding the fundamental processes of life.

\subsection*{Acknowledgments}
P.R. was funded by the Deutsche Forschungsgemeinschaft (DFG) (German Research Foundation), grant RO47766/2-1 and acknowledges funding by the DFG under Germany’s Excellence Strategy–EXC 2002/1 “Science of Intelligence,” project 390523135. B.C.D. was supported by a fellowship at the Wissenschaftskolleg zu Berlin and by the ASU-SFI Center for Biosocial Complex Systems. 

\clearpage
\bibliographystyle{ws-rv-van}
\bibliography{phase_trans_coll_behav}

\providecommand{\noopsort}[1]{}\providecommand{\singleletter}[1]{#1}%
\begin{thebibliography}{143}
\providecommand{\natexlab}[1]{#1}
\providecommand{\url}[1]{\texttt{#1}}
\expandafter\ifx\csname urlstyle\endcsname\relax
  \providecommand{\doi}[1]{doi: #1}\else
  \providecommand{\doi}{doi: \begingroup \urlstyle{rm}\Url}\fi

\bibitem{boccara2010modeling}
N.~Boccara, \emph{Modeling complex systems}. Springer Science \& Business Media
   (2010).

\bibitem{newman2011complex}
M.~E. Newman, Complex systems: A survey, \emph{arXiv preprint arXiv:1112.1440}
  (2011).

\bibitem{kwapien2012physical}
J.~Kwapie{\'n} and S.~Dro{\.z}d{\.z}, Physical approach to complex systems,
  \emph{Physics Reports}. {\bf 515}\penalty0 (3-4), \penalty0 115--226  (2012).

\bibitem{ising1924beitrag}
E.~Ising.
\newblock \emph{Beitrag zur Theorie des Ferro- und Paramagnetismus}.
\newblock PhD thesis, Grefe \& Tiedemann  (1924).

\bibitem{newell1953theory}
G.~F. Newell and E.~W. Montroll, On the theory of the ising model of
  ferromagnetism, \emph{Reviews of Modern Physics}. {\bf 25}\penalty0 (2),
  \penalty0 353  (1953).

\bibitem{ising2017fate}
T.~Ising, R.~Folk, R.~Kenna, B.~Berche, and Y.~Holovatch, The fate of
  \uppercase{E}rnst \uppercase{I}sing and the fate of his model., \emph{Journal
  of Physical Studies}. {\bf 21}\penalty0 (3), \penalty0 3002  (2017).

\bibitem{stanley1971phase}
H.~E. Stanley, \emph{Phase transitions and critical phenomena}. vol.~7,
  Clarendon Press, Oxford  (1971).

\bibitem{domb2001phase}
C.~Domb and J.~Lebowitz, eds., \emph{Phase transitions and critical phenomena}.
  vol.~19, Academic Press (Elsevier)  (2001).

\bibitem{hinrichsen2006non}
H.~Hinrichsen, Non-equilibrium phase transitions, \emph{Physica A: Statistical
  Mechanics and its Applications}. {\bf 369}\penalty0 (1), \penalty0 1--28
  (2006).

\bibitem{davidson2006gene}
E.~H. Davidson and D.~H. Erwin, Gene regulatory networks and the evolution of
  animal body plans, \emph{Science}. {\bf 311}\penalty0 (5762), \penalty0
  796--800  (2006).

\bibitem{beggs2012being}
J.~M. Beggs and N.~Timme, Being critical of criticality in the brain,
  \emph{Frontiers in Physiology}. {\bf 3}, \penalty0 163  (2012).

\bibitem{kinouchi2006optimal}
O.~Kinouchi and M.~Copelli, Optimal dynamical range of excitable networks at
  criticality, \emph{Nature Physics}. {\bf 2}\penalty0 (5), \penalty0 348--351
  (2006).

\bibitem{moussaid2009collective}
M.~Moussaid, S.~Garnier, G.~Theraulaz, and D.~Helbing, Collective information
  processing and pattern formation in swarms, flocks, and crowds, \emph{Topics
  in Cognitive Science}. {\bf 1}\penalty0 (3), \penalty0 469--497  (2009).

\bibitem{munoz2018colloquium}
M.~A. Munoz, Colloquium: Criticality and dynamical scaling in living systems,
  \emph{Reviews of Modern Physics}. {\bf 90}\penalty0 (3), \penalty0 031001
  (2018).

\bibitem{bak1987self}
P.~Bak, C.~Tang, and K.~Wiesenfeld, Self-organized criticality: An explanation
  of the 1/f noise, \emph{Physical Review Letters}. {\bf 59}\penalty0 (4),
  \penalty0 381  (1987).

\bibitem{bak1988self}
P.~Bak, C.~Tang, and K.~Wiesenfeld, Self-organized criticality, \emph{Physical
  Review A}. {\bf 38}\penalty0 (1), \penalty0 364  (1988).

\bibitem{markovic2014power}
D.~Markovi{\'c} and C.~Gros, Power laws and self-organized criticality in
  theory and nature, \emph{Physics Reports}. {\bf 536}\penalty0 (2), \penalty0
  41--74  (2014).

\bibitem{jensen1998self}
H.~J. Jensen, \emph{Self-organized criticality: emergent complex behavior in
  physical and biological systems}. vol.~10, Cambridge university press
  (1998).

\bibitem{turcotte1999self}
D.~L. Turcotte, Self-organized criticality, \emph{Reports on Progress in
  Physics}. {\bf 62}\penalty0 (10), \penalty0 1377  (1999).

\bibitem{beggs2003neuronal}
J.~M. Beggs and D.~Plenz, Neuronal avalanches in neocortical circuits,
  \emph{Journal of Neuroscience}. {\bf 23}\penalty0 (35), \penalty0
  11167--11177  (2003).

\bibitem{hesse2014self}
J.~Hesse and T.~Gross, Self-organized criticality as a fundamental property of
  neural systems, \emph{Frontiers in Systems Neuroscience}. {\bf 8}, \penalty0
  166  (2014).

\bibitem{priesemann2014spike}
V.~Priesemann, M.~Wibral, M.~Valderrama, R.~Pr{\"o}pper, M.~Le~Van~Quyen,
  T.~Geisel, J.~Triesch, D.~Nikoli{\'c}, and M.~H. Munk, Spike avalanches in
  vivo suggest a driven, slightly subcritical brain state, \emph{Frontiers in
  Systems Neuroscience}. {\bf 8}, \penalty0 108  (2014).

\bibitem{plenz2021self}
D.~Plenz, T.~L. Ribeiro, S.~R. Miller, P.~A. Kells, A.~Vakili, and E.~L. Capek,
  Self-organized criticality in the brain, \emph{Frontiers in Physics}. {\bf
  9}, \penalty0 365  (2021).

\bibitem{balleza2008critical}
E.~Balleza, E.~R. Alvarez-Buylla, A.~Chaos, S.~Kauffman, I.~Shmulevich, and
  M.~Aldana, Critical dynamics in genetic regulatory networks: examples from
  four kingdoms, \emph{PLoS One}. {\bf 3}\penalty0 (6), \penalty0 e2456
  (2008).

\bibitem{daniels2018criticality}
B.~C. Daniels, H.~Kim, D.~Moore, S.~Zhou, H.~B. Smith, B.~Karas, S.~A.
  Kauffman, and S.~I. Walker, Criticality distinguishes the ensemble of
  biological regulatory networks, \emph{Physical Review Letters}. {\bf
  121}\penalty0 (13), \penalty0 138102  (2018).

\bibitem{mora2011biological}
T.~Mora and W.~Bialek, Are biological systems poised at criticality?,
  \emph{Journal of Statistical Physics}. {\bf 144}\penalty0 (2), \penalty0
  268--302  (2011).

\bibitem{calovi2015collective}
D.~S. Calovi, U.~Lopez, P.~Schuhmacher, H.~Chat{\'e}, C.~Sire, and
  G.~Theraulaz, Collective response to perturbations in a data-driven fish
  school model, \emph{Journal of The Royal Society Interface}. {\bf
  12}\penalty0 (104), \penalty0 20141362  (2015).

\bibitem{klamser2021collective}
P.~P. Klamser and P.~Romanczuk, Collective predator evasion: Putting the
  criticality hypothesis to the test, \emph{PLoS Computational Biology}. {\bf
  17}\penalty0 (3), \penalty0 e1008832  (2021).

\bibitem{poel2022subcritical}
W.~Poel, B.~C. Daniels, M.~M.~G. Sosna, C.~R. Twomey, S.~P. Leblanc, I.~D.
  Couzin, and P.~Romanczuk, Subcritical escape waves in schooling fish,
  \emph{Science Advances}. {\bf 8}\penalty0 (25), \penalty0 eabm6385  (2022).
\newblock \doi{10.1126/sciadv.abm6385}.

\bibitem{bierbach2017behavioural}
D.~Bierbach, K.~L. Laskowski, and M.~Wolf, Behavioural individuality in clonal
  fish arises despite near-identical rearing conditions, \emph{Nature
  Communications}. {\bf 8}\penalty0 (1), \penalty0 1--7  (2017).

\bibitem{wild2021social}
B.~Wild, D.~M. Dormagen, A.~Zachariae, M.~L. Smith, K.~S. Traynor,
  D.~Brockmann, I.~D. Couzin, and T.~Landgraf, Social networks predict the life
  and death of honey bees, \emph{Nature Communications}. {\bf 12}\penalty0 (1),
  \penalty0 1--12  (2021).

\bibitem{bonabeau1997self}
E.~Bonabeau, G.~Theraulaz, J.-L. Deneubourg, S.~Aron, and S.~Camazine,
  Self-organization in social insects, \emph{Trends in Ecology \& Evolution}.
  {\bf 12}\penalty0 (5), \penalty0 188--193  (1997).

\bibitem{gordon2019ecology}
D.~M. Gordon et~al., The ecology of collective behavior in ants, \emph{Annual
  Review of Entomology}. {\bf 64}, \penalty0 35--50  (2019).

\bibitem{strandburg2015shared}
A.~Strandburg-Peshkin, D.~R. Farine, I.~D. Couzin, and M.~C. Crofoot, Shared
  decision-making drives collective movement in wild baboons, \emph{Science}.
  {\bf 348}\penalty0 (6241), \penalty0 1358--1361  (2015).

\bibitem{marro2005nonequilibrium}
J.~Marro and R.~Dickman, Nonequilibrium phase transitions in lattice models,
  \emph{Nonequilibrium Phase Transitions in Lattice Models}  (2005).

\bibitem{henkel2008non}
M.~Henkel, H.~Hinrichsen, S.~L{\"u}beck, and M.~Pleimling,
  \emph{Non-equilibrium phase transitions}. vol.~1, Springer  (2008).

\bibitem{vicsek1995novel}
T.~Vicsek, A.~Czir{\'o}k, E.~Ben-Jacob, I.~Cohen, and O.~Shochet, Novel type of
  phase transition in a system of self-driven particles, \emph{Physical Review
  Letters}. {\bf 75}\penalty0 (6), \penalty0 1226  (1995).

\bibitem{BuhSumCou06}
J.~Buhl, D.~J. Sumpter, I.~D. Couzin, J.~Hale, E.~Despland, E.~Miller, and
  S.~Simpson, {From Disorder to Order in Marching Locusts}, \emph{Science}.
  {\bf 312}\penalty0 (June), \penalty0 1402--1406  (2006).

\bibitem{szabo2006phase}
B.~Szabo, G.~Sz{\"o}ll{\"o}si, B.~G{\"o}nci, Z.~Jur{\'a}nyi, D.~Selmeczi, and
  T.~Vicsek, Phase transition in the collective migration of tissue cells:
  experiment and model, \emph{Physical Review E}. {\bf 74}\penalty0 (6),
  \penalty0 061908  (2006).

\bibitem{FeiPinGel18}
O.~Feinerman, I.~Pinkoviezky, A.~Gelblum, E.~Fonio, and N.~S. Gov, {The physics
  of cooperative transport by ants}, \emph{Nature Physics}. pp. 1--31  (2018).
\newblock ISSN 1745-2473.
\newblock \doi{10.1038/s41567-018-0107-y}.

\bibitem{misund2003schooling}
O.~A. Misund, J.~Coetzee, P.~Fr{\'e}on, M.~Gardener, K.~Olsen, I.~Svellingen,
  and I.~Hampton, Schooling behaviour of sardine sardinops sagax in false bay,
  south africa, \emph{African Journal of Marine Science}. {\bf 25}, \penalty0
  185--193  (2003).

\bibitem{cavagna2013boundary}
A.~Cavagna, I.~Giardina, and F.~Ginelli, Boundary information inflow enhances
  correlation in flocking, \emph{Physical Review Letters}. {\bf 110}\penalty0
  (16), \penalty0 168107  (2013).

\bibitem{davidson2021collective}
J.~D. Davidson, M.~M. Sosna, C.~R. Twomey, V.~H. Sridhar, S.~P. Leblanc, and
  I.~D. Couzin, Collective detection based on visual information in animal
  groups, \emph{Journal of the Royal Society Interface}. {\bf 18}\penalty0
  (180), \penalty0 20210142  (2021).

\bibitem{ballerini2008interaction}
M.~Ballerini, N.~Cabibbo, R.~Candelier, A.~Cavagna, E.~Cisbani, I.~Giardina,
  V.~Lecomte, A.~Orlandi, G.~Parisi, A.~Procaccini, et~al., Interaction ruling
  animal collective behavior depends on topological rather than metric
  distance: Evidence from a field study, \emph{Proceedings of the National
  Academy of Sciences}. {\bf 105}\penalty0 (4), \penalty0 1232--1237  (2008).

\bibitem{mermin1966absence}
N.~D. Mermin and H.~Wagner, Absence of ferromagnetism or antiferromagnetism in
  one- or two-dimensional isotropic heisenberg models, \emph{Physical Review
  Letters}. {\bf 17}, \penalty0 1133--1136  (Nov, 1966).
\newblock \doi{10.1103/PhysRevLett.17.1133}.

\bibitem{hohenberg1967existance}
P.~C. Hohenberg, Existence of long-range order in one and two dimensions,
  \emph{Physical Review}. {\bf 158}, \penalty0 383--386  (Jun, 1967).
\newblock \doi{10.1103/PhysRev.158.383}.

\bibitem{toner1995long}
J.~Toner and Y.~Tu, Long-range order in a two-dimensional dynamical xy model:
  how birds fly together, \emph{Physical Review Letters}. {\bf 75}\penalty0
  (23), \penalty0 4326  (1995).

\bibitem{toner1998flocks}
J.~Toner and Y.~Tu, Flocks, herds, and schools: A quantitative theory of
  flocking, \emph{Physical Review E}. {\bf 58}\penalty0 (4), \penalty0 4828
  (1998).

\bibitem{toner2005hydrodynamics}
J.~Toner, Y.~Tu, and S.~Ramaswamy, Hydrodynamics and phases of flocks,
  \emph{Annals of Physics}. {\bf 318}\penalty0 (1), \penalty0 170--244  (2005).

\bibitem{gregoire2004onset}
G.~Gr{\'e}goire and H.~Chat{\'e}, Onset of collective and cohesive motion,
  \emph{Physical Review Letters}. {\bf 92}\penalty0 (2), \penalty0 025702
  (2004).

\bibitem{chate2008modeling}
H.~Chat{\'e}, F.~Ginelli, G.~Gr{\'e}goire, F.~Peruani, and F.~Raynaud, Modeling
  collective motion: variations on the vicsek model, \emph{The European
  Physical Journal B}. {\bf 64}\penalty0 (3), \penalty0 451--456  (2008).

\bibitem{baglietto2009nature}
G.~Baglietto and E.~V. Albano, Nature of the order-disorder transition in the
  vicsek model for the collective motion of self-propelled particles,
  \emph{Physical Review E}. {\bf 80}\penalty0 (5), \penalty0 050103  (2009).

\bibitem{ginelli2016physics}
F.~Ginelli, The physics of the vicsek model, \emph{The European Physical
  Journal Special Topics}. {\bf 225}\penalty0 (11), \penalty0 2099--2117
  (2016).

\bibitem{chate2006simple}
H.~Chat{\'e}, F.~Ginelli, and R.~Montagne, Simple model for active nematics:
  Quasi-long-range order and giant fluctuations, \emph{Physical Review
  Letters}. {\bf 96}\penalty0 (18), \penalty0 180602  (2006).

\bibitem{peruani2011polar}
F.~Peruani, F.~Ginelli, M.~B{\"a}r, and H.~Chat{\'e}.
\newblock Polar vs. apolar alignment in systems of polar self-propelled
  particles.
\newblock In \emph{Journal of Physics: Conference Series}, vol. 297, p. 012014
  (2011).

\bibitem{baskaran2012self}
A.~Baskaran and M.~C. Marchetti, Self-regulation in self-propelled nematic
  fluids, \emph{The European Physical Journal E}. {\bf 35}\penalty0 (9),
  \penalty0 1--8  (2012).

\bibitem{giomi2012banding}
L.~Giomi, L.~Mahadevan, B.~Chakraborty, and M.~Hagan, Banding, excitability and
  chaos in active nematic suspensions, \emph{Nonlinearity}. {\bf 25}\penalty0
  (8), \penalty0 2245  (2012).

\bibitem{bertin2015comparison}
E.~Bertin, A.~Baskaran, H.~Chat{\'e}, and M.~C. Marchetti, Comparison between
  smoluchowski and boltzmann approaches for self-propelled rods, \emph{Physical
  Review E}. {\bf 92}\penalty0 (4), \penalty0 042141  (2015).

\bibitem{grossmann2016mesoscale}
R.~Gro{\ss}mann, F.~Peruani, and M.~B{\"a}r, Mesoscale pattern formation of
  self-propelled rods with velocity reversal, \emph{Physical Review E}. {\bf
  94}\penalty0 (5), \penalty0 050602  (2016).

\bibitem{bertin2006boltzmann}
E.~Bertin, M.~Droz, and G.~Gr{\'e}goire, Boltzmann and hydrodynamic description
  for self-propelled particles, \emph{Physical Review E}. {\bf 74}\penalty0
  (2), \penalty0 022101  (2006).

\bibitem{ihle2011kinetic}
T.~Ihle, Kinetic theory of flocking: Derivation of hydrodynamic equations,
  \emph{Physical Review E}. {\bf 83}\penalty0 (3), \penalty0 030901  (2011).

\bibitem{ihle2013invasion}
T.~Ihle, Invasion-wave-induced first-order phase transition in systems of
  active particles, \emph{Physical Review E}. {\bf 88}\penalty0 (4), \penalty0
  040303  (2013).

\bibitem{chate2008collective}
H.~Chat{\'e}, F.~Ginelli, G.~Gr{\'e}goire, and F.~Raynaud, Collective motion of
  self-propelled particles interacting without cohesion, \emph{Physical Review
  E}. {\bf 77}\penalty0 (4), \penalty0 046113  (2008).

\bibitem{bertin2009hydrodynamic}
E.~Bertin, M.~Droz, and G.~Gr{\'e}goire, Hydrodynamic equations for
  self-propelled particles: microscopic derivation and stability analysis,
  \emph{Journal of Physics A: Mathematical and Theoretical}. {\bf 42}\penalty0
  (44), \penalty0 445001  (2009).

\bibitem{grossman2008emergence}
D.~Grossman, I.~Aranson, and E.~B. Jacob, Emergence of agent swarm migration
  and vortex formation through inelastic collisions, \emph{New Journal of
  Physics}. {\bf 10}\penalty0 (2), \penalty0 023036  (2008).

\bibitem{hanke2013understanding}
T.~Hanke, C.~A. Weber, and E.~Frey, Understanding collective dynamics of soft
  active colloids by binary scattering, \emph{Physical Review E}. {\bf
  88}\penalty0 (5), \penalty0 052309  (2013).

\bibitem{grossmann2013self}
R.~Grossmann, L.~Schimansky-Geier, and P.~Romanczuk, Self-propelled particles
  with selective attraction--repulsion interaction: from microscopic dynamics
  to coarse-grained theories, \emph{New Journal of Physics}. {\bf 15}\penalty0
  (8), \penalty0 085014  (2013).

\bibitem{romanczuk2009collective}
P.~Romanczuk, I.~D. Couzin, and L.~Schimansky-Geier, Collective motion due to
  individual escape and pursuit response, \emph{Physical Review Letters}. {\bf
  102}\penalty0 (1), \penalty0 010602  (2009).

\bibitem{ferrante2013elasticity}
E.~Ferrante, A.~E. Turgut, M.~Dorigo, and C.~Huepe, Elasticity-based mechanism
  for the collective motion of self-propelled particles with springlike
  interactions: a model system for natural and artificial swarms,
  \emph{Physical Review Letters}. {\bf 111}\penalty0 (26), \penalty0 268302
  (2013).

\bibitem{ginelli2010relevance}
F.~Ginelli and H.~Chat{\'e}, Relevance of metric-free interactions in flocking
  phenomena, \emph{Physical Review Letters}. {\bf 105}\penalty0 (16), \penalty0
  168103  (2010).

\bibitem{rahmani2021topological}
P.~Rahmani, F.~Peruani, and P.~Romanczuk, Topological flocking models in
  spatially heterogeneous environments, \emph{Communications Physics}. {\bf
  4}\penalty0 (1), \penalty0 1--9  (2021).

\bibitem{martin2021fluctuation}
D.~Martin, H.~Chat{\'e}, C.~Nardini, A.~Solon, J.~Tailleur, and F.~Van~Wijland,
  Fluctuation-induced phase separation in metric and topological models of
  collective motion, \emph{Physical Review Letters}. {\bf 126}\penalty0 (14),
  \penalty0 148001  (2021).

\bibitem{grossmann2012active}
R.~Grossmann, L.~Schimansky-Geier, and P.~Romanczuk, Active brownian particles
  with velocity-alignment and active fluctuations, \emph{New Journal of
  Physics}. {\bf 14}\penalty0 (7), \penalty0 073033  (2012).

\bibitem{klamser2021impact}
P.~Klamser, L.~G{\'o}mez-Nava, T.~Landgraf, J.~Jolles, D.~Bierbach, and
  P.~Romanczuk, Impact of variable speed on collective movement of animal
  groups, \emph{Frontiers in Physics}. {\bf 9}  (2021).

\bibitem{cavagna2022marginal}
A.~Cavagna, A.~Culla, X.~Feng, I.~Giardina, T.~S. Grigera, W.~Kion-Crosby,
  S.~Melillo, G.~Pisegna, L.~Postiglione, and P.~Villegas, Marginal speed
  confinement resolves the conflict between correlation and control in
  collective behaviour, \emph{Nature Communications}. {\bf 13}\penalty0 (1),
  \penalty0 1--11  (2022).

\bibitem{BiaCavGia14}
W.~Bialek, A.~Cavagna, I.~Giardina, T.~Mora, O.~Pohl, E.~Silvestri, M.~Viale,
  and A.~Walczac, {Social interactions dominate speed control in poising
  natural flocks near criticality}, \emph{Proceedings of the National Academy
  of Sciences, USA}. {\bf 111}\penalty0 (20), \penalty0 7212  (2014).

\bibitem{attanasi2014information}
A.~Attanasi, A.~Cavagna, L.~Del~Castello, I.~Giardina, T.~S. Grigera,
  A.~Jeli{\'c}, S.~Melillo, L.~Parisi, O.~Pohl, E.~Shen, et~al., Information
  transfer and behavioural inertia in starling flocks, \emph{Nature Physics}.
  {\bf 10}\penalty0 (9), \penalty0 691--696  (2014).

\bibitem{cavagna2015flocking}
A.~Cavagna, L.~Del~Castello, I.~Giardina, T.~Grigera, A.~Jelic, S.~Melillo,
  T.~Mora, L.~Parisi, E.~Silvestri, M.~Viale, et~al., Flocking and turning: a
  new model for self-organized collective motion, \emph{Journal of Statistical
  Physics}. {\bf 158}\penalty0 (3), \penalty0 601--627  (2015).

\bibitem{cavagna2019dynamical}
A.~Cavagna, L.~Di~Carlo, I.~Giardina, L.~Grandinetti, T.~S. Grigera, and
  G.~Pisegna, Dynamical renormalization group approach to the collective
  behavior of swarms, \emph{Physical Review Letters}. {\bf 123}\penalty0 (26),
  \penalty0 268001  (2019).

\bibitem{visscher2007group}
P.~K. Visscher, Group decision making in nest-site selection among social
  insects, \emph{Annual Review of Entomology}. {\bf 52}, \penalty0 255--275
  (2007).

\bibitem{couzin2011uninformed}
I.~D. Couzin, C.~C. Ioannou, G.~Demirel, T.~Gross, C.~J. Torney, A.~Hartnett,
  L.~Conradt, S.~A. Levin, and N.~E. Leonard, Uninformed individuals promote
  democratic consensus in animal groups, \emph{Science}. {\bf 334}\penalty0
  (6062), \penalty0 1578--1580  (2011).

\bibitem{DanFlaKra17}
B.~C. Daniels, J.~C. Flack, and D.~C. Krakauer, Dual coding theory explains
  biphasic collective computation in neural decision-making, \emph{Frontiers in
  Neuroscience}. {\bf 11}, \penalty0 1--16  (2017).
\newblock ISSN 1662-453X.
\newblock \doi{10.3389/fnins.2017.00313}.

\bibitem{BruKraFla18}
E.~R. Brush, D.~C. Krakauer, and J.~C. Flack, Conflicts of interest improve
  collective computation of adaptive social structures, \emph{Science
  Advances}. {\bf 4}, \penalty0 1--11  (2018).
\newblock ISSN 23752548.
\newblock \doi{10.1126/sciadv.1603311}.

\bibitem{arganda2012common}
S.~Arganda, A.~P{\'e}rez-Escudero, and G.~G. de~Polavieja, A common rule for
  decision making in animal collectives across species, \emph{Proceedings of
  the National Academy of Sciences}. {\bf 109}\penalty0 (50), \penalty0
  20508--20513  (2012).

\bibitem{pais2013mechanism}
D.~Pais, P.~M. Hogan, T.~Schlegel, N.~R. Franks, N.~E. Leonard, and J.~A.
  Marshall, A mechanism for value-sensitive decision-making, \emph{PloS One}.
  {\bf 8}\penalty0 (9), \penalty0 e73216  (2013).

\bibitem{turalska2009complexity}
M.~Turalska, M.~Lukovic, B.~J. West, and P.~Grigolini, Complexity and
  synchronization, \emph{Physical Review E}. {\bf 80}\penalty0 (2), \penalty0
  021110  (2009).

\bibitem{hartnett2016heterogeneous}
A.~T. Hartnett, E.~Schertzer, S.~A. Levin, and I.~D. Couzin, Heterogeneous
  preference and local nonlinearity in consensus decision making,
  \emph{Physical Review Letters}. {\bf 116}\penalty0 (3), \penalty0 038701
  (2016).

\bibitem{pinkoviezky2018collective}
I.~Pinkoviezky, I.~D. Couzin, and N.~S. Gov, Collective conflict resolution in
  groups on the move, \emph{Physical Review E}. {\bf 97}\penalty0 (3),
  \penalty0 032304  (2018).

\bibitem{lee2014simple}
C.~H. Lee and A.~Lucas, Simple model for multiple-choice collective decision
  making, \emph{Physical Review E}. {\bf 90}\penalty0 (5), \penalty0 052804
  (2014).

\bibitem{castellano2009statistical}
C.~Castellano, S.~Fortunato, and V.~Loreto, Statistical physics of social
  dynamics, \emph{Reviews of Modern Physics}. {\bf 81}\penalty0 (2), \penalty0
  591  (2009).

\bibitem{srivastava2014collective}
V.~Srivastava and N.~E. Leonard, Collective decision-making in ideal networks:
  The speed-accuracy tradeoff, \emph{IEEE Transactions on Control of Network
  Systems}. {\bf 1}\penalty0 (1), \penalty0 121--132  (2014).

\bibitem{tump2020wise}
A.~N. Tump, T.~J. Pleskac, and R.~H. Kurvers, Wise or mad crowds? the cognitive
  mechanisms underlying information cascades, \emph{Science Advances}. {\bf
  6}\penalty0 (29), \penalty0 eabb0266  (2020).

\bibitem{gronlund2007dynamic}
A.~Gr{\"o}nlund, P.~Holme, and P.~Minnhagen, Dynamic scaling regimes of
  collective decision making, \emph{EPL (Europhysics Letters)}. {\bf
  81}\penalty0 (2), \penalty0 28003  (2007).

\bibitem{winklmayr2020wisdom}
C.~Winklmayr, A.~B. Kao, J.~B. Bak-Coleman, and P.~Romanczuk, The wisdom of
  stalemates: consensus and clustering as filtering mechanisms for improving
  collective accuracy, \emph{Proceedings of the Royal Society B}. {\bf
  287}\penalty0 (1938), \penalty0 20201802  (2020).

\bibitem{couzin2005effective}
I.~D. Couzin, J.~Krause, N.~R. Franks, and S.~A. Levin, Effective leadership
  and decision-making in animal groups on the move, \emph{Nature}. {\bf
  433}\penalty0 (7025), \penalty0 513--516  (2005).

\bibitem{GelPinFon15}
A.~Gelblum, I.~Pinkoviezky, E.~Fonio, A.~Ghosh, N.~Gov, and O.~Feinerman, {Ant
  groups optimally amplify the effect of transiently informed individuals},
  \emph{Nature Communications}. {\bf 6}, \penalty0 7729  (2015).
\newblock ISSN 2041-1723.
\newblock \doi{10.1038/ncomms8729}.

\bibitem{DanRom21}
B.~C. Daniels and P.~Romanczuk, {Quantifying the impact of network structure on
  speed and accuracy in collective decision-making}, \emph{Theory in
  Biosciences}. {\bf 140}, \penalty0 379--390  (2021).
\newblock ISSN 23318422.
\newblock \doi{10.1007/s12064-020-00335-1}.

\bibitem{bose2019bifurcation}
I.~Bose and S.~Ghosh, Bifurcation and criticality, \emph{Journal of Statistical
  Mechanics: Theory and Experiment}. {\bf 2019}\penalty0 (4), \penalty0 043403
  (2019).

\bibitem{gross2021not}
T.~Gross, Not one, but many critical states: a dynamical systems perspective,
  \emph{Frontiers in Neural Circuits}. {\bf 15}, \penalty0 614268  (2021).

\bibitem{dodds2004universal}
P.~S. Dodds and D.~J. Watts, Universal behavior in a generalized model of
  contagion, \emph{Physical Review Letters}. {\bf 92}\penalty0 (21), \penalty0
  218701  (2004).

\bibitem{bottcher2017critical}
L.~B{\"o}ttcher, J.~Nagler, and H.~J. Herrmann, Critical behaviors in contagion
  dynamics, \emph{Physical Review Letters}. {\bf 118}\penalty0 (8), \penalty0
  088301  (2017).

\bibitem{sosna2019individual}
M.~M. Sosna, C.~R. Twomey, J.~Bak-Coleman, W.~Poel, B.~C. Daniels,
  P.~Romanczuk, and I.~D. Couzin, Individual and collective encoding of risk in
  animal groups, \emph{Proceedings of the National Academy of Sciences}. {\bf
  116}\penalty0 (41), \penalty0 20556--20561  (2019).

\bibitem{DanKraFla17}
B.~C. Daniels, D.~C. Krakauer, and J.~C. Flack, {Control of finite critical
  behaviour in a small-scale social system}, \emph{Nature Communications}. {\bf
  8}, \penalty0 14301  (2017).
\newblock ISSN 2041-1723.
\newblock \doi{10.1038/ncomms14301}.

\bibitem{araujo2014recent}
N.~Ara{\'u}jo, P.~Grassberger, B.~Kahng, K.~Schrenk, and R.~M. Ziff, Recent
  advances and open challenges in percolation, \emph{The European Physical
  Journal Special Topics}. {\bf 223}\penalty0 (11), \penalty0 2307--2321
  (2014).

\bibitem{granovetter1978threshold}
M.~Granovetter, Threshold models of collective behavior, \emph{American Journal
  of Sociology}. {\bf 83}\penalty0 (6), \penalty0 1420--1443  (1978).

\bibitem{rosenthal2015revealing}
S.~B. Rosenthal, C.~R. Twomey, A.~T. Hartnett, H.~S. Wu, and I.~D. Couzin,
  Revealing the hidden networks of interaction in mobile animal groups allows
  prediction of complex behavioral contagion, \emph{Proceedings of the National
  Academy of Sciences}. {\bf 112}\penalty0 (15), \penalty0 4690--4695  (2015).

\bibitem{ruan2015kinetics}
Z.~Ruan, G.~Iniguez, M.~Karsai, and J.~Kert{\'e}sz, Kinetics of social
  contagion, \emph{Physical Review Letters}. {\bf 115}\penalty0 (21), \penalty0
  218702  (2015).

\bibitem{iacopini2019simplicial}
I.~Iacopini, G.~Petri, A.~Barrat, and V.~Latora, Simplicial models of social
  contagion, \emph{Nature Communications}. {\bf 10}\penalty0 (1), \penalty0
  1--9  (2019).

\bibitem{dodds2005generalized}
P.~S. Dodds and D.~J. Watts, A generalized model of social and biological
  contagion, \emph{Journal of Theoretical Biology}. {\bf 232}\penalty0 (4),
  \penalty0 587--604  (2005).

\bibitem{SarHayPel21}
R.~Sarfati, J.~C. Hayes, and O.~Peleg, {Self-organization in natural swarms of
  Photinus carolinus synchronous fireflies}, \emph{Science Advances}. {\bf
  7}\penalty0 (28), \penalty0 1--6  (2021).
\newblock ISSN 23752548.
\newblock \doi{10.1126/sciadv.abg9259}.

\bibitem{Kur84}
Y.~Kuramoto, \emph{{Chemical Oscillations, Waves, and Turbulence}}.
  Springer-Verlag, New York  (1984).

\bibitem{AceBonVic05}
J.~A. Acebr{\'{o}}n, L.~L. Bonilla, C.~J. Vicente, F.~Ritort, and R.~Spigler,
  {The Kuramoto model: A simple paradigm for synchronization phenomena},
  \emph{Reviews of Modern Physics}. {\bf 77}\penalty0 (1), \penalty0 137--185
  (2005).
\newblock ISSN 00346861.
\newblock \doi{10.1103/RevModPhys.77.137}.

\bibitem{MirStr90}
R.~E. Mirollo and S.~H. Strogatz, {Synchronization of Pulse-Coupled Biological
  Oscillators}, \emph{SIAM Journal on Applied Mathematics}. {\bf 50}\penalty0
  (6), \penalty0 1645--1662  (1990).

\bibitem{DanEllKra16}
B.~C. Daniels, C.~J. Ellison, D.~C. Krakauer, and J.~C. Flack, {Quantifying
  collectivity}, \emph{Current Opinion in Neurobiology}. {\bf 37}, \penalty0
  106--113  (2016).
\newblock \doi{http://dx.doi.org/10.1016/j.conb.2016.01.012}.

\bibitem{ProLizObs11}
M.~Prokopenko, J.~T. Lizier, O.~Obst, and X.~R. Wang, {Relating Fisher
  information to order parameters}, \emph{Physical Review E}. {\bf 84}\penalty0
  (4), \penalty0 041116  (oct, 2011).
\newblock ISSN 1539-3755.
\newblock \doi{10.1103/PhysRevE.84.041116}.

\bibitem{vynnycky2010introduction}
E.~Vynnycky and R.~White, \emph{An introduction to infectious disease
  modelling}. OUP oxford  (2010).

\bibitem{SheYanPet09}
W.~L. Shew, H.~Yang, T.~Petermann, R.~Roy, and D.~Plenz, {Neuronal avalanches
  imply maximum dynamic range in cortical networks at criticality.}, \emph{The
  Journal of Neuroscience}. {\bf 29}\penalty0 (49), \penalty0 15595--600  (dec,
  2009).
\newblock ISSN 1529-2401.
\newblock \doi{10.1523/JNEUROSCI.3864-09.2009}.

\bibitem{HeiRosHag15}
A.~M. Hein, S.~B. Rosenthal, G.~I. Hagstrom, A.~Berdahl, C.~J. Torney, and
  I.~D. Couzin, {The evolution of distributed sensing and collective
  computation in animal populations}, \emph{eLife}. {\bf 4}, \penalty0 e10955
  (2015).
\newblock ISSN 2050084X.
\newblock \doi{10.7554/eLife.10955}.

\bibitem{langton1990computation}
C.~G. Langton, Computation at the edge of chaos: Phase transitions and emergent
  computation, \emph{Physica D: Nonlinear phenomena}. {\bf 42}\penalty0 (1-3),
  \penalty0 12--37  (1990).

\bibitem{teuscher2022revisiting}
C.~Teuscher, Revisiting the edge of chaos: Again?, \emph{Biosystems}. p. 104693
   (2022).

\bibitem{cramer2020control}
B.~Cramer, D.~St{\"o}ckel, M.~Kreft, M.~Wibral, J.~Schemmel, K.~Meier, and
  V.~Priesemann, Control of criticality and computation in spiking neuromorphic
  networks with plasticity, \emph{Nature Communications}. {\bf 11}\penalty0
  (1), \penalty0 1--11  (2020).

\bibitem{vanni2011criticality}
F.~Vanni, M.~Lukovi{\'c}, and P.~Grigolini, Criticality and transmission of
  information in a swarm of cooperative units, \emph{Physical Review Letters}.
  {\bf 107}\penalty0 (7), \penalty0 078103  (2011).

\bibitem{franks2003speed}
N.~R. Franks, A.~Dornhaus, J.~P. Fitzsimmons, and M.~Stevens, Speed versus
  accuracy in collective decision making, \emph{Proceedings of the Royal
  Society of London. Series B: Biological Sciences}. {\bf 270}\penalty0 (1532),
  \penalty0 2457--2463  (2003).

\bibitem{marshall2009optimal}
J.~a~R~Marshall, R.~Bogacz, A.~Dornhaus, R.~Planqué, T.~Kovacs, and N.~R.
  Franks, On optimal decision-making in brains and social insect colonies.,
  \emph{Journal of the Royal Society, Interface}. {\bf 6}, \penalty0 1065--74
  (11, 2009).
\newblock ISSN 1742-5662.
\newblock \doi{10.1098/rsif.2008.0511}.

\bibitem{AreJinDan18}
E.~Arehart, T.~Jin, and B.~C. Daniels, {Locating Decision-Making Circuits in a
  Heterogeneous Neural Network}, \emph{Frontiers in Applied Mathematics and
  Statistics}. {\bf 4}, \penalty0 11  (2018).
\newblock \doi{10.3389/fams.2018.00011}.

\bibitem{LewCra08}
S.~M. Lewis and C.~K. Cratsley, {Flash signal evolution, mate choice, and
  predation in fireflies}, \emph{Annual Review of Entomology}. {\bf 53},
  \penalty0 293--321  (2008).
\newblock ISSN 00664170.
\newblock \doi{10.1146/annurev.ento.53.103106.093346}.

\bibitem{bornholdt2000topological}
S.~Bornholdt and T.~Rohlf, Topological evolution of dynamical networks: Global
  criticality from local dynamics, \emph{Physical Review Letters}. {\bf
  84}\penalty0 (26), \penalty0 6114  (2000).

\bibitem{attanasi2014finite}
A.~Attanasi, A.~Cavagna, L.~Del~Castello, I.~Giardina, S.~Melillo, L.~Parisi,
  O.~Pohl, B.~Rossaro, E.~Shen, E.~Silvestri, et~al., Finite-size scaling as a
  way to probe near-criticality in natural swarms, \emph{Physical Review
  Letters}. {\bf 113}\penalty0 (23), \penalty0 238102  (2014).

\bibitem{hidalgo2014information}
J.~Hidalgo, J.~Grilli, S.~Suweis, M.~A. Munoz, J.~R. Banavar, and A.~Maritan,
  Information-based fitness and the emergence of criticality in living systems,
  \emph{Proceedings of the National Academy of Sciences}. {\bf 111}\penalty0
  (28), \penalty0 10095--10100  (2014).

\bibitem{brush2016content}
E.~R. Brush, N.~E. Leonard, and S.~A. Levin, The content and availability of
  information affects the evolution of social-information gathering strategies,
  \emph{Theoretical Ecology}. {\bf 9}\penalty0 (4), \penalty0 455--476  (2016).

\bibitem{moretti2013griffiths}
P.~Moretti and M.~A. Munoz, Griffiths phases and the stretching of criticality
  in brain networks, \emph{Nature Communications}. {\bf 4}\penalty0 (1),
  \penalty0 1--10  (2013).

\bibitem{o2008good}
R.~O'Gorman, K.~M. Sheldon, and D.~S. Wilson, For the good of the group?
  exploring group-level evolutionary adaptations using multilevel selection
  theory., \emph{Group Dynamics: Theory, Research, and Practice}. {\bf
  12}\penalty0 (1), \penalty0 17  (2008).

\bibitem{cooney2019replicator}
D.~B. Cooney, The replicator dynamics for multilevel selection in evolutionary
  games, \emph{Journal of Mathematical Biology}. {\bf 79}\penalty0 (1),
  \penalty0 101--154  (2019).

\bibitem{bucsoniu2010multi}
L.~Bu{\c{s}}oniu, R.~Babu{\v{s}}ka, and B.~D. Schutter, Multi-agent
  reinforcement learning: An overview, \emph{Innovations in multi-agent systems
  and applications-1}. pp. 183--221  (2010).

\bibitem{romero2019idtracker}
F.~Romero-Ferrero, M.~G. Bergomi, R.~C. Hinz, F.~J. Heras, and G.~G.
  De~Polavieja, Idtracker. ai: tracking all individuals in small or large
  collectives of unmarked animals, \emph{Nature Methods}. {\bf 16}\penalty0
  (2), \penalty0 179--182  (2019).

\bibitem{walter2021trex}
T.~Walter and I.~D. Couzin, Trex, a fast multi-animal tracking system with
  markerless identification, and 2d estimation of posture and visual fields,
  \emph{eLife}. {\bf 10}, \penalty0 e64000  (2021).

\bibitem{Dan20}
B.~C. Daniels.
\newblock {Inferring the Logic of Collective Information Processors}.
\newblock In eds. M.~Chen, J.~M. Dunn, A.~Golan, and A.~Ullah, \emph{Advances
  in Info-Metrics: Information and Information Processing across Disciplines}.
  Oxford University Press  (2020).
\newblock ISBN 9780190636685.
\newblock \doi{10.1093/oso/9780190636685.003.0003}.

\bibitem{SchNemMeh14}
D.~J. Schwab, I.~Nemenman, and P.~Mehta, {Zipf's Law and Criticality in
  Multivariate Data without Fine-Tuning.}, \emph{Physical Review Letters}. {\bf
  113}\penalty0 (6), \penalty0 068102  (aug, 2014).
\newblock ISSN 1079-7114.

\bibitem{nurse2021biology}
P.~Nurse et~al., Biology must generate ideas as well as data, \emph{Nature}.
  {\bf 597}\penalty0 (7876), \penalty0 305--305  (2021).

\bibitem{lima1995back}
S.~L. Lima, Back to the basics of anti-predatory vigilance: the group-size
  effect, \emph{Animal Behaviour}. {\bf 49}\penalty0 (1), \penalty0 11--20
  (1995).

\bibitem{gordon2014ecology}
D.~M. Gordon, The ecology of collective behavior, \emph{PLoS biology}. {\bf
  12}\penalty0 (3), \penalty0 e1001805  (2014).

\bibitem{monk2018ecology}
C.~T. Monk, M.~Barbier, P.~Romanczuk, J.~R. Watson, J.~Al{\'o}s, S.~Nakayama,
  D.~I. Rubenstein, S.~A. Levin, and R.~Arlinghaus, How ecology shapes
  exploitation: a framework to predict the behavioural response of human and
  animal foragers along exploration--exploitation trade-offs, \emph{Ecology
  letters}. {\bf 21}\penalty0 (6), \penalty0 779--793  (2018).

\bibitem{davis2022using}
G.~H. Davis, M.~C. Crofoot, and D.~R. Farine, Using optimal foraging theory to
  infer how groups make collective decisions, \emph{Trends in Ecology \&
  Evolution}  (2022).

\end{thebibliography}

\printindex                        

\end{document}